 \documentclass[preprintnumbers,amsmath,amssymb,floatfix]{revtex4}

\usepackage{graphicx}
\usepackage{dcolumn}
\usepackage{bm}
\usepackage{amssymb}
\usepackage{epsfig}
\usepackage{color}

\newcommand{\ba}{\begin{eqnarray}}
\newcommand{\ea}{\end{eqnarray}}
\newcommand{\be}{\begin{equation}}
\newcommand{\ee}{\end{equation}}
\newcommand{\bdisplay}{\begin{displaymath}}
\newcommand{\edisplay}{\end{displaymath}}

\newcommand{\eq}[1]{Eq.\,(\ref{#1})}

\newcommand{\la}{\,\raisebox{-.8ex}{\,$\stackrel{\textstyle <}{\sim}$}\,\,} 
\newcommand{\ga}{\,\raisebox{-.8ex}{\,$\stackrel{\textstyle >}{\sim}$}\,\,}

\begin{document}

\title{An analytic solution to  LO coupled  DGLAP evolution equations: a new pQCD tool}
\author{Martin~M.~Block}
\affiliation{Department of Physics and Astronomy, Northwestern University, 
Evanston, IL 60208}
\author{Loyal Durand}
\affiliation{Department of Physics, University of Wisconsin, Madison, WI 53706}
\author{Phuoc Ha}
\affiliation{Department of Physics, Astronomy and Geosciences, Towson University , Towson, MD 21252}
\author{Douglas W. McKay}
\affiliation{Department of Physics and Astronomy, University of Kansas, Lawrence, KS 66045} 
\date{\today}

\begin{abstract}
We have analytically solved the LO pQCD (leading order perturbative QCD) singlet DGLAP (Dokshitzer, Gribov, Lipatov, Alterelli, Parisi) equations \cite{dglap1,dglap2,dglap3} using Laplace transform techniques. 
Newly-developed highly accurate numerical inverse Laplace transform algorithms \cite{inverseLaplace1,inverseLaplace2} 
 allow us to write  fully decoupled solutions for the singlet structure function $F_s(x,Q^2)$ and $G(x,Q^2)$ as   
\ba
F_s(x,Q^2)={\cal F}_s(F_{s0}(x_0), G_0(x_0))\quad {\rm and} \quad
G(x,Q^2)={\cal G}(F_{s0}(x_0), G_0(x_0)),\nonumber
\ea 
where the $x_0$ are the Bjorken-$x$ values at $Q_0^2$. 
Here ${\cal F}_s$ and $\cal G$ are known functions---found using LO DGLAP splitting functions---of the initial boundary conditions $F_{s0}(x)\equiv F_s(x,Q_0^2)$ and $G_{0}(x)\equiv G(x,Q_0^2)$, i.e.,  the chosen starting functions at the virtuality $Q_0^2$. For both $G(x)$ and $F_s(x)$,  we are able to {\em either devolve or evolve} each separately and rapidly, with very high numerical accuracy, a computational fractional precision of $O(10^{-9})$. Armed with this powerful new tool in the pQCD arsenal, we  compare our numerical results from the above equations with the published MSTW2008  and CTEQ6L LO gluon and singlet $F_s$  distributions \cite{MSTW1}, starting from their initial  values at $Q_0^2=1$ GeV$^2$ and $1.69$  GeV$^2$, respectively, using their choices of $\alpha_s(Q^2)$. This allows an important independent check on the accuracies of their evolution codes and therefore  the computational accuracies of their published parton distributions.  Our method completely decouples the two LO distributions, at the same time  guaranteeing that both $G$ and $F_s$  satisfy the singlet coupled DGLAP equations.  It also allows one to  easily obtain the effects of the starting functions  on the evolved gluon and singlet structure functions, as functions of both $Q^2$ and $Q_0^2$, being equally accurate in devolution ($Q^2<Q_0^2$) as in evolution ($Q^2>Q_0^2$). Further, it can also be used for non-singlet distributions, thus giving LO analytic  solutions for individual quark and gluon distributions at a given $x$ and $Q^2$, rather than the numerical solutions of the coupled  integral-differential equations on a large, but fixed, two-dimensional grid that are currently available.
\end{abstract}


\maketitle


\section{Introduction} \label{sec:introduction} 
The search for new physics at the LHC demands an accurate knowledge of gluon distribution functions at small Bjorken $x$ and large virtuality $Q^2$, both for estimating QCD backgrounds and for calculating gluon-initiated processes.
The traditional method has simultaneously determined gluon and quark distribution functions  by fitting experimental data on neutral- and charged-current deep inelastic scattering processes and some jet data over a large domain  of values of $x$ and $Q^2$.  The distributions at small $x$ and large $Q^2$ are determined mainly by the  proton structure function $F_2^{\gamma p}(x,Q^2)$ measured in deep inelastic $ep$ (or $\gamma^*p$) scattering.  The fitting process starts with an initial  $Q^2_0$, typically less than or equal to the square of the  $c$ quark mass, $m_c^2\approx 2$ GeV$^2$, and  individual quark and gluon trial distributions parameterized with pre-determined shapes,  given as functions of $x$ for the chosen $Q_0^2$. The distributions are then evolved numerically on a finite, albeit large, two-dimensional grid in $x$ and $Q^2$  to larger $Q^2$ using the coupled integral-differential DGLAP equations  \cite{dglap1,dglap2,dglap3}, typically in leading order (LO) and next-to- leading order (NLO), and the results used to predict  measured quantities. The final distributions are then determined by adjusting the input parameters  to obtain a best fit to experimental  data, fitting both HERA and Tevatron data over a large range of $x$ and $Q^2$,  along with selected hard scattering data from fixed target experiments. This procedure is very indirect in the case of the gluon: the gluon distribution $G(x,Q^2) = xg(x,Q^2)$ does not contribute directly to the accurately determined structure function $F_2^{\gamma p}(x,Q^2)$, and is determined only through the quark distributions in conjunction with the evolution equations, or at large $x$, from jet data. For recent determinations of the gluon and quark distributions, see \cite{CTEQ6.1,CTEQ6L,CTEQ6.5,MRST,MRST4, MSTW1}.

In the following, we will summarize our method for  analytically determining $G(x,Q^2)$ and the singlet structure function  $F_s(x,Q^2)$ {\em directly } and {\em individually}, using as input $G_0(x)\equiv G(x,Q^2_0)$ and $F_{s0}(x)\equiv F_s(x,Q^2_0)$, where $Q_0^2$ is arbitrary, with the guarantee that each distribution  individually satisfies the coupled DGLAP equations.   The method is readily  extended  to embrace non-singlet functions, so that it can be used also to find individual quark distributions.  However, we will not pursue that goal in this communication. Instead, we give a numerical demonstration which takes advantage of the fact that our analytic solutions achieve numerical accuracies of $O(10^{-9})$, giving us a new diagnostic tool to verify {\em published} LO singlet structure functions ($F_s(x,Q^2)$)  and gluon ($G(x,Q^2)=xg(x,Q^2$)) distributions. In order to test the numerical accuracy of their evolution codes, we consider two cases,   using the published LO starting distributions for  $G_0$ and  $F_{s0}$:
\begin{enumerate}
\item MSTW2008 \cite{MSTW1}:  for $Q_0^2=1$ GeV$^2$,  we generate LO singlet structure functions and gluon distributions  \cite{MSTW1}, using the strong coupling constant $\alpha_s(Q^2)$ that they used for their LO evolution and compare them with their published values MSTW2008 \cite{MSTW1} for the domain $10^{-6}\le x \le 1$ and $1\le Q^2\le 100000$. We find that their evolution code has serious  problems at small $x$, producing significant numerical inaccuracies.  It should be noted that the MSTW group {\em does not} do devolution.
 \item \label{item:cteq} CTEQ6L \cite{CTEQ6.1}: for  $Q_0^2=1.69$ GeV$^2$, we generate LO singlet structure functions and gluon distributions, using the strong coupling constant $\alpha_s(Q^2)$  \cite{CTEQ6L} they used for both LO evolution and devolution. With our high numerical precision at {\em all} $x$ and $Q^2$,  we are able to verify all of their published {\em evolution} results---to larger $Q^2$---but show that their published {\em devolution} results, i.e.,  $Q^2<Q^2_0$, have significant numerical inaccuracies at small $x$. 
\end{enumerate} 
Finally, using our accurate CTEQ devolution results, we compare  LO starting distributions  for both groups at $Q^2=1$ GeV$^2$, noting that the CTEQ6L LO gluon distribution {\em turns over} and  {\em  goes negative} at small $x$, i.e., $x\la 5\times10^{-5}$, whereas the MSTW2008 LO gluon starting distribution {\em continues to rise sharply} at small $x$. 

\section{Decoupling the coupled LO singlet DGLAP equations} 
Our approach uses  a somewhat unusual application of Laplace transforms \cite{bdm1,bdm2}, in which we first introduce the variable $v\equiv \ln(1/x)$ into the coupled DGLAP equations,  then Laplace transform  these coupled integral-differential equations in $v$ space to obtain coupled {\em } homogeneous first-order differential equations in the Laplace-space variable $s$. We  solve these equations analytically. Finally, using fast and accurate numerical inverse Laplace transform algorithms \cite{inverseLaplace1,inverseLaplace2}, we transform the  solutions back into $v$ space, and, finally, into Bjorken $x$-space, so that we can write 
\ba F_s(x,Q^2)={\cal F}_s(F_{s0}(x_0), G_0(x_0))\quad  {\rm and} \quad
G(x,Q^2)={\cal G}(F_{s0}(x_0), G_0(x_0)),
\ea
 where the functions $\cal F$ and $\cal G$ are determined by the splitting functions in the DGLAP equations, with $x_0$ being the Bjorken-$x$ at the starting virtuality $Q_0^2$;  $F_{s0}(x)$ and $G_0(x)$ are the known starting distributions at $Q^2=Q_0^2$, where evolution (devolution) begins.

Our method can be generalized to NLO (see Ref. \cite{bdhmNLO}), but for brevity, we will limit ourselves to LO in this paper.  We write  the coupled LO DGLAP equations  \cite{bdm1,bdm2}  as
\ba
\frac{4\pi}{\alpha_s(Q^2)}\frac{\partial F_s}{\partial\ln{Q^2}}(x,Q^2)&=& 4{F_s(x,Q^2)}+\frac{16}{3}{F_s(x,Q^2)}\ln\frac{1-x}{x}
+\frac{16}{3}x\int_x^1\left(\frac{F_s(z,Q^2)}{z}-\frac{F_s(x,Q^2)}{x}\right)\frac{dz}{ z-x}\nonumber\\
&&-\frac{8}{3}x\int_x^1F_s(z,Q^2)\left(1+\frac{x}{z}\right)\frac{\,dz}{z^2}+2n_fx\int_x^1G(z,Q^2)\left(1-2{x\over z}+2{x^2\over z^2}\right)\,{dz\over z^2},\label{dFdtauofx}\\
\frac{4\pi}{\alpha_s(Q^2)}\frac{\partial G}{\partial \ln{Q^2}}(x,Q^2)&=& {33-2n_f\over 3}{G(x,Q^2)}+12{G(x,Q^2)}\ln\frac{1-x}{x}
+12x\int_x^1\left(\frac{G(z,Q^2)}{z}-\frac{G(x,Q^2)}{x}\right)\frac{dz}{ z-x}\nonumber\\
&&+12x\int_x^1G(z,Q^2)\left({z\over x} -2 +{x\over z} -{x^2\over z^2}\right)\frac{\,dz}{z^2}+{8\over 3}\int_x^1 F_s(z,Q^2)\left(1+\left(1-{x\over z}\right)^2    \right)\,{dz\over z}.\label{dGdtauofx}
\ea
Here $\alpha_s(Q^2)$ is the running strong coupling constant,  and for LO MSTW2008 \cite{MSTW1} is given  by the LO form
\be
\alpha_s(Q^2)={4\pi \over \left( 11-{2\over 3}n_f   \right)\ln(Q^2/\Lambda_{n_f}^2)}\label{alphasMSTW},
\ee
with $n_f$ the number of quark flavors. The QCD parameter $\Lambda_{5}$ is fixed so that the known  $\alpha_s(M_Z^2)$  is reproduced and then $\Lambda_4$ and $\Lambda_3$ are adjusted so that $\alpha_s$ is continuous across the boundaries $Q^2=M_b^2$ and $M_c^2$, respectively, where $M_b$ and $M_c$ are the masses of the $b$ and $c$ quarks. Later, we will also introduce the NLO form of $\alpha_s$ used (with $\alpha_s(M_Z^2)=0.118)$ in their LO CTEQ6L \cite{CTEQ6.1} evolution, when we discuss CTEQ6L pdfs.
 
We now examine the last two terms of line 1 in \eq{dFdtauofx} and rewrite them, introducing the variable changes $v=\ln (1/x)$, $w=\ln (1/z)$,  and the notation $\hat F_s(v,Q^2)\equiv F_s(e^{-v},Q^2)$,  $\hat G(v,Q^2 )\equiv G(e^{-v},Q^2)$, as
\ba
&&\frac{16}{3}{\hat F_s(v,Q^2)}\ln (e^v-1)
+\frac{16}{3}\int_0^v\left(\hat F_s(w,Q^2)-\hat F_s(v,Q^2)e^{v-w}\right)\frac{1}{e^{v-w}-1}\,dw\nonumber\\
&&=\frac{16}{3}\int_0^v{\partial \hat F_s\over\partial w}(w,Q^2)\ln \left(1-e^{-(v-w)}\right)\,dw.\label{rewrite}
\ea
where the final result---the last line in \eq{rewrite}---is found by replacing the upper limit $v$ in integral of line 1 of \eq{rewrite} by $v-\epsilon$, carrying out the integrals, doing a partial integration  and finally, taking the limit as $\epsilon \rightarrow 0$. Similarly, we find for the last two terms of line 1 in \eq{dGdtauofx}, that
\ba
&&12{\hat G(v,Q^2)}\ln (e^v-1)
+12\int_0^v\left(\hat G(w,Q^2)-\hat G(v,Q^2)e^{v-w}\right)\frac{1}{e^{v-w}-1}\,dw\nonumber\\
&&=12\int_0^v{\partial\hat G \over\partial w}(w,Q^2)\ln \left(1-e^{-(v-w)}\right)\,dw.\label{rewriteG}
\ea

We now rewrite \eq{dFdtauofx} and \eq{dGdtauofx} in terms of the new variable $v$ as
\ba
\frac{4\pi}{\alpha_s(Q^2)}\frac{\partial \hat F_s}{\partial \ln{Q^2}}(v,Q^2)&=& 4{\hat F_s(v,Q^2)}
+\frac{16}{3}\int_0^v{\partial \hat F_s\over\partial w}(w,Q^2)\ln \left(1-e^{w-v}\right)\,dw\nonumber\\
&&-\frac{8}{3}\int_0^v\hat F_s(w,Q^2)\left( e^{-(v-w)}+e^{-2(v-w)}   \right)\,dw\nonumber\\
&&+2n_fx\int_0^v\hat G(w,Q^2)\left(e^{-(v-w)} -2e^{-2(v-w)}+2e^{-3(v-w)}   \right)\,dw,\label{dFdtauofv}\\
\frac{4\pi}{\alpha_s(Q^2)}\frac{\partial \hat G}{\partial \ln{Q^2}}(v,Q^2)&=& {33-2n_f\over 3}{\hat G(v,Q^2)}
+12\int_0^v{\partial\hat G \over\partial w}(w,Q^2)\ln \left(1-e^{-(v-w)}\right)\,dw\nonumber\\
&&+12\int_0^v \hat G(w,Q^2)\left(1-2e^{-(v-w)}+e^{-2(v-w)}-e^{-3(v-w)} \right) \,dw\nonumber\\
&&+{8\over 3}\int_0^v\hat F_s(w,Q^2)\left( 1+\left(1-e^{-(v-w)}\right)^2            \right)\,dw.\label{dGdtauofv}
\ea
The DGLAP equations have now been written in a form such that all of the integrals in \eq{dFdtauofv} and \eq{dGdtauofv} are  manifestly seen to be  convolution integrals. Thus, introducing Laplace transforms  allows us to factor these convolution integrals, since the Laplace transform of a convolution is the product of the Laplace transforms of the factors, i.e.,
\ba
{\cal L}\left[\int_0^v F[w]H[v-w]\,dw;s   \right]&=&{\cal L}\left[\int_0^v F[v-w]H[w]\,dw;s   \right]={\cal L} [F[v];s]\times {\cal L} [H[v];s]\label{convolution}.
\ea
Defining the Laplace transforms of $\hat F_s(v,Q^2)$ and $\hat G(v,Q^2)$ in $s$ space as
\ba
f(s,Q^2)&\equiv &{\cal L}\left[ \hat F_s(v,Q^2);s\right]=\int^\infty_0{\hat F_s}(v,Q^2)e^{-sv}\,dv,\quad
g(s,Q^2)\equiv {\cal L}[\hat G(v,Q^2);s]=\int^\infty_0{\hat G}(v,Q^2)e^{-sv}\,dv
\ea
and noting that 
\ba
{\cal L}\left[{\partial \hat F_s \over\partial w}(w,Q^2);s\right]=s f(s,Q^2),\qquad
{\cal L}\left[{\partial \hat G \over\partial w}(w,Q^2);s\right]=s g(s,Q^2),
\ea
since $ F_s (v=0,Q^2)=G(v=0,Q^2)=0$,
we now factor the Laplace transforms of \eq{dFdtauofv} and \eq{dGdtauofv} into two coupled  first order differential equations in Laplace space $s$ having $Q^2$-dependent coefficients. These can be written as
\ba
{\partial f\over \partial \ln{Q^2}}(s,Q^2) &=&\frac{\alpha_s(Q^2)}{4\pi}\Phi_f (s)f(s,Q^2)+\frac{\alpha_s(Q^2)}{4\pi}\Theta_f(s)g(s,Q^2)\label{df},\\
{\partial g\over \partial \ln{Q^2}}(s,Q^2) &=&\frac{\alpha_s(Q^2)}{4\pi}\Phi_g (s)g(s,Q^2)+\frac{\alpha_s(Q^2)}{4\pi}\Theta_g(s)f(s,Q^2).\label{dg}
\ea 
The coefficient functions $\Phi$ and $\Theta$ are given by 
\ba
\Phi_f(s)&=&4 -{8\over 3}\left({1\over s+1}+{1\over s+2}+2\left(\psi(s+1)+\gamma_E\right)\right)\label{Phif}\\
\Theta_f(s)&=&2n_f\left({1\over s+1}-{2\over s+2}+{2\over s+3} \right),\label{Thetaf}\\
\Phi_g(s)&=&{33-2n_f \over 3} +12\left({1\over s}-{ 2\over s+1}+{1\over s+2}-{1 \over s+3}-\psi(s+1)-\gamma_E\right)\label{Phig}\\
\Theta_g(s)&=&{8\over 3}\left({2\over s}-{2\over s+1}+{1\over s+2}     \right),\label{Thetag}
\ea
where $\psi(x)$ is the digamma function and $\gamma_E=0.5772156\ldots$ is Euler's constant.

The solution of the coupled equations  in \eq{df} and \eq{dg} in terms of initial values of the functions $f$ and $g$, specified as functions of $s$ at virtuality $Q_0^2$, is straightforward. The $Q^2$ dependence of the solutions is expressed entirely through the function 
\be
\tau(Q^2,Q_0^2)={1\over 4 \pi}\int_{Q_0^2}^{Q^2} \alpha_s(Q'^2)\,d\,\ln Q'^2\label{tau}.
\ee
With the initial conditions $f_0(s)\equiv f(s,Q_0^2)$ and $g_0(s)\equiv g(s,Q_0^2)$, the solutions are
\ba
f(s,\tau)&= &k_{ff}(s,\tau)f_0(s)+k_{fg}(s,\tau) g_0(s),\label{f}\\
g(s,\tau)&= &k_{gg}(s,\tau)g_0(s)+k_{gf}(s,\tau) f_0(s)\label{g},
\ea 
where the coefficient functions in the solution are
\ba
k_{ff}(s,\tau)&\equiv&e^{\frac{{\tau }}{2}\left(\Phi_f(s) +\Phi_g(s)\right)}\left[\cosh\left (  {\tau \over 2}R(s)\right) +\frac{\sinh\left({\tau\over2}R(s)\right)}{R(s)} \left(\Phi_f(s)-\Phi_g(s)\right)\right],\label{kff}\\
k_{fg}(s,\tau)&\equiv &e^{ {\tau\over 2}\left(\Phi_f(s)+\Phi_g(s)\right) }{2\sinh\left ( {\tau \over 2}R(s) \right )\over R(s)}\,\Theta_f(s),\label{kfg}\\
k_{gg}(s,\tau)&\equiv &e^{{\tau \over2}\left(\Phi_f(s) +\Phi_g(s)\right)}\left[\cosh\left (  {\tau \over 2}R(s)\right) -\frac{\sinh\left({\tau\over2}R(s)\right)}{R(s)} \left(\Phi_f(s)-\Phi_g(s)\right)\right],\label{kgg}\\
k_{gf}(s,\tau)&\equiv&e^{ {\tau\over 2}\left(\Phi_f(s)+\Phi_g(s)\right) }{2\sinh\left ( {\tau \over 2}R(s) \right )\over R(s)}\,\Theta_g(s),\label{kgf}
\ea
with $R(s) \equiv \sqrt{\left(\Phi_f(s)-\Phi_g(s)\right)^2+4 \Theta_f(s)\Theta_g(s)}$. Clearly, the fundamental solutions in Laplace space $s$, \eq{f} and \eq{g}, are symmetric under the interchange $f\leftrightarrow g$.

Let us now define four  kernels $K_{FF},\ K_{FG},\ K_{GF}$ and $K_{GG}$, the inverse Laplace transforms of the $k's$, i.e.,
\ba
K_{FF}(v,\tau) &\equiv &{\cal L}^{-1}[k_{ff}(s,\tau);v],\qquad K_{FG}(v,\tau) \equiv {\cal L}^{-1}[k_{fg}(s,\tau);v],\label{fKernels}\\
K_{GG}(v,\tau) &\equiv &{\cal L}^{-1}[k_{gg}(s,\tau);v],\qquad K_{GF}(v,\tau) \equiv {\cal L}^{-1}[k_{gf}(s,\tau);v].\label{gKernels}
\ea
It is evident from Eqs.\ (\ref{tau}),  (\ref{kfg}), and (\ref{kgf}) that $K_{FG}$ and $K_{GF}$ vanish for $Q^2=Q_0^2$ where $\tau(Q^2,Q_0^2)=0$. It can also be shown without difficulty that for $\tau=0$, $K_{FF}(v,0)=K_{GG}(v,0)=\delta(v)$ and that $K_{FG}(v,0)=K_{GF}(v,0)=0$.

The initial boundary conditions at $Q_0^2$ are given by $F_{s0}(x)=F_s(x,Q^2_0)$ and $G_0(x)=G(x,Q^2_0)$. 
In $v$-space,
$\hat F_{s0}(v)\equiv F_{s0}(e^{-v})$ and $\hat G_0(v)\equiv G_0(e^{-v})$
 are the inverse Laplace transforms of $f_{0}(s)$ and $g_0(s)$, respectively, i.e.,
\ba
\hat F_{s0}(v)&\equiv &{\cal L}^{-1}[f_0(s);v]\ {\rm and \ } \hat G_0(v)\equiv {\cal L}^{-1}[g_0(s);v].
\ea

Finally, we can write  our  {\em decoupled} singlet structure function $\hat F_s$ and $\hat G$ solutions  in $v$-space in terms of the convolution integrals as
\ba
\hat F_s(v,Q^2)&=&\int_0^v K_{FF}(v-w,\tau(Q^2,Q_0^2))\hat F_{s0}(w)\,dw +\int_0^v K_{FG}(v-w,\tau(Q^2,Q_0^2))\hat G_0(w)\,dw, \label{F}\\
\hat G(v,Q^2)&=&\int_0^v K_{GG}(v-w,\tau(Q^2,Q_0^2))\hat G_0(w)\,dw +\int_0^v K_{GF}(v-w,\tau(Q^2,Q_0^2))\hat F_{s0}(w)\,dw .\label{G}
\ea

We now derive an alternate form of the solution to the decoupled equation, very useful for computational purposes, that does not use the convolution theorem. Using a suitable fast and accurate numerical inverse Laplace transform \cite{inverseLaplace1}, we can {\em directly} invert \eq{f} and \eq{g}, since $f_0(s)$ and $g_0(s)$---the  Laplace transforms of the {\em known} starting functions $\hat F_{s0}(v,Q^2)$
 and $\hat G_{0}(v,Q^2)$---are readily obtainable; the coefficient functions,  the $k$'s given in Eq.(\ref{kff}--\ref{kgf}), are known functions of $s$ and $\tau$, and hence, of $Q^2$ and $Q_0^2$.  Thus we finally  write our decoupled analytic solution in $v$ space as 
\ba
\hat F_s(v,Q^2)&=&{\cal L}^{-1}[\left(k_{ff}(s,\tau)f_0(s)+k_{fg}(s,\tau) g_0(s)\right);v],\label{newF}\\
\hat G(v,Q^2)&=&{\cal L}^{-1}[\left(k_{gg}(s,\tau)g_0(s)+k_{gf}(s,\tau) f_0(s)\right);v]\label{newG}.
\ea

In order to use our solution in the integral representation of \eq{F} and \eq{G}, we must first {\em numerically} invert Laplace transforms of the type of $k_{ff}$ and $k_{gg}$ that for small $\tau$ look similar to Dirac $\delta$ functions; a formidable numerical task that is inherently inaccurate, and is thus computationally intensive and significantly  slower (but possible)  using the numerical inverse transforms of Ref. \cite{inverseLaplace2}.  On the other hand, if we use   \eq{newF} and \eq{newG}, we only have to invert a function  whose inverse Laplace transform ($\hat F_s(v,Q^2)$ or $\hat G(v,Q^2)$) is very smooth and thus can be well approximated by a high order polynomial in $v$. As shown in Ref. \cite{inverseLaplace2}, it can then in principle be evaluated to arbitrary accuracy very rapidly. It will be  shown  in the Appendix that we actually achieve a fractional  accuracy of $O(10^{-11})$ in our numerical Laplace inversion.   In Section \ref{section:accuracy} we will do a  detailed evaluation of the inherent {\em overall numerical  accuracy} for actual physical problems, showing that we can do {\em both devolution and evolution} rapidly to fractional accuracies of $O(10^{-9})$ using the numerical methods outlined in the Appendix.

The final desired {\em decoupled}   $F_s(x,Q^2)$ and $G(x,Q^2)$ in Bjorken-$x$ space 
 are readily found by substituting  $v=\ln(1/x)$ into the $v$-space solutions for $\hat F_s(v,Q^2)$ and $\hat G(v,Q^2)$ from 
\eq{newF} and \eq{newG}.

\section{Analytic LO non-singlet distributions}
For {\em non-singlet} distributions $F_{ns}(x,Q^2)$, such as the difference between the $u$ and $d$ quark distributions,  $x\left[u(x,Q^2)-d(x,Q^2)\right]$, we can schematically write the logarithmic derivative of $F_{ns}$ as the convolution of $F_{ns}(x,Q^2)$ with the non-singlet splitting function ${\cal K}_{ns}(x)$ (using the convolution symbol $\otimes$), i.e., 
\ba
{4\pi\over \alpha_s(Q^2)}{\partial F_{ns}\over \partial \ln (Q^2)}(x,Q^2)&=&F_{ns}\otimes {{\cal K}}_{ns}.
\ea
After again changing to the variable $v=\ln(1/x)$ and going to Laplace space $s$, we find the simple solution
\ba
f_{ns}(s,\tau)&=&e^{\tau \Phi_{ns}(s)}f_{ns0}(s),\qquad{\rm where\ } \Phi_{ns}(s)={\cal L}\left [e^{-v}\hat{\cal K}_{ns}(v);s\right ]\quad {\rm and\ }\hat{\cal K}_{ns}(v) ={\cal K}_{ns}\left(e^{-v}\right).\label{NSofs}
\ea
Thus we can find {\em any} non-singlet solution in $v$-space, using the non-singlet kernel $K_{ns}(v)\equiv {\cal L}^{-1}\left[e^{\tau \Phi_{ns}(s)};v   \right]$, by either employing  the Laplace  convolution relation  
\ba
F_{ns}(v,Q^2)=\int _0^v K_{ns}(v-w,\tau(Q^2,Q_0^2))\hat F_{ns0}(w)\,dw\label{Fnsofv}
\ea
or  the non-integral form
\ba
F_{ns}(v,Q^2)={\cal L}^{-1}\left[e^{\tau \Phi_{ns}(s)}f_{ns0}(s);v   \right]\label{newFnsofv}.
\ea
 In this case, either method works equally well numerically, since the non-singlet functions $K_{ns}(v)$ can also be approximated by a polynomial in $v$.

For brevity, we will not pursue the case of the non-singlet solution any further here except to note that in LO the $\Phi_{ns}(s)$  in \eq{NSofs} is identical to $\Phi_{f}(s)$ defined in  \eq{Phif}. Instead, we will concentrate on the more difficult case of $F_s$ and $G$. 
\section{ LO MSTW2008 singlet and gluon distributions}\label{section:singletMSTW}
As an example of the application of our analytic decoupled solutions, we will use  the published MSTW2008 initial starting functions $F_{s0}(x)$ and $G_0(x)$ at $Q_0^2=1$ GeV$^2$ \cite{MSTW1} and  will compare  our LO $x$-space gluon distribution $G(x,Q^2)=xg(x,Q^2)$ using \eq{newG} and our LO singlet structure function $F_s(x,Q^2)$  using  \eq{newF}---both numerically evaluated using a powerful new inverse Laplace transformation algorithm \cite {inverseLaplace1}---with the  corresponding LO  distributions published by the MSTW collaboration \cite{MSTW1}.
In order to insure continuity across the boundaries $Q^2=M_c^2$ and $M_b^2$, we will first evolve from $Q_0^2=1$ GeV$^2$ (the MSTW $Q_0^2$ value) to $M_c^2$ and use our evolved values of $G(x,M_c^2)$ and $F_s(x,M_c^2)$ for {\em new} starting values $G_0(x)$ and $F_{s0}(x)$. We will then evolve to $M_b^2$, repeating the process, thus insuring continuity of $G$ and $F_s$ at the boundaries where $n_f$ changes. We use the MSTW values  $M_c=1.40$ GeV, $M_b=4.75$ GeV, $\alpha_s(1\ {\rm GeV}^2)=0.6818$  and  $\alpha_s(M_Z^2)=0.13939$ in their definition of $\alpha_s(Q^2)$ in \eq{alphasMSTW}.

\subsection{$G(x,Q^2)$ and $Fs(x,Q^2)$ for LO MSTW2008}
\begin{figure}[h]
\begin{center}
\mbox{\epsfig{file=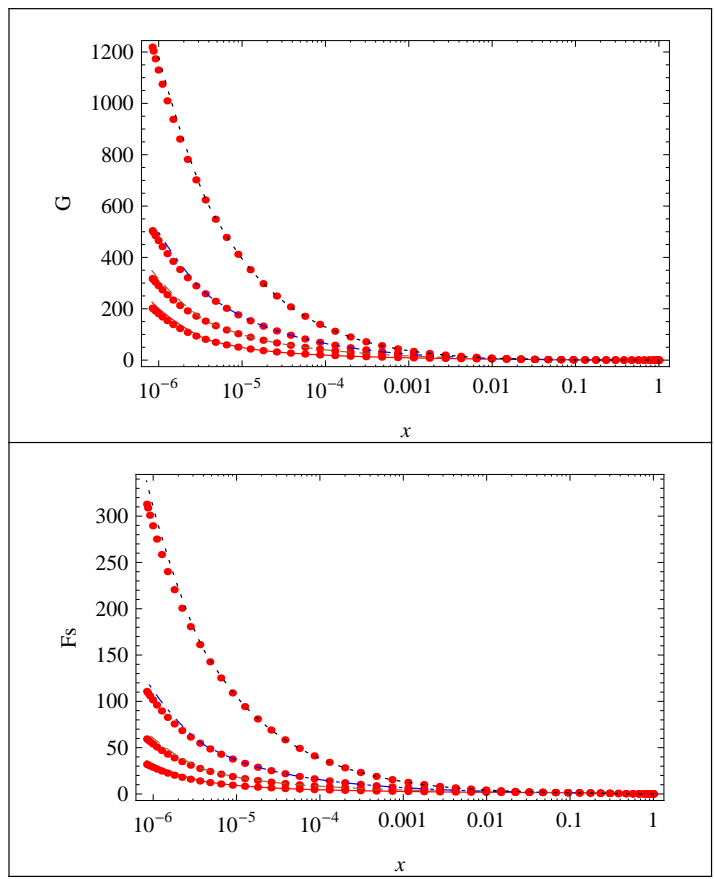
,width=4in%
,bbllx=0pt,bblly=0pt,bburx=317pt,bbury=370pt,clip=%
}}
\end{center}
\caption[]{
Plots for LO MSTW2008 \cite{MSTW1} gluon distributions $G(x,Q^2)=xg(x,Q^2)$ (upper plot) and $F_s(x,Q^2)$ distributions (lower plot) vs. Bjorken $x$.
 The MSTW2008 curves are for $Q^2 = 5,\ 20,\ 100$ and $M_Z^2$ GeV$^2$, bottom  to top. The (red) dots are our evolution results for LO $G(x,Q^2)$ from \eq{newG} and $F_s$ from \eq{newF}, after converting  to $x$-space, using the LO MSTW2008 \cite{MSTW1} values for $F_{s0}(x)$ and G$_0(x)$, with their choice of  $Q^2_0=1$ GeV$^2$. The $x$ range covers all of the published LO MSTW2008 $x$ data.
} 
\label{fig:GFs}
\end{figure}

In Fig. \ref{fig:GFs} we show the LO  $x$-space results for  $G(x,Q^2)=xg(x,Q^2)$ (upper figure) and  $F_s(x,Q^2)$ (lower figure) vs. $x$, for 4 representative values of $Q^2$. The $x$-domain, $10^{-6}\le x \le 1$, is the complete region covered by the MSTW group \cite{MSTW1}.   The curves are the published LO MSTW2008    distributions \cite{MSTW1}: from bottom  to top; the (red) curve is for $Q^2 = 5$ GeV$^2$; the (brown) dashed curve is for $Q^2 = 20$ GeV$^2$;  the (blue) dot-dashed curve is for $Q^2 = 100$ GeV$^2$;   the (black) dotted curve is for $Q^2 =M_Z^2$. The (red) dots are our analytic results for LO $G(x,Q^2)$ from \eq{newG} and $F_s(x,Q^2)$ from \eq{newF}, converted to $x$-space, using the LO MSTW2008 values for $F_{s0}(x)$ and G$_0(x)$; the numerical values were evaluated using {\em Mathematica} \cite{Mathematica}. An outline of the numerical procedure is given in the Appendix.

For large $x$, the agreement  is excellent for all $Q^2$. However, as seen in a close inspection of Fig. \ref{fig:GFs}, the disagreement for both $G$ and $F_s$ becomes significantly large as we go to small $x$. We will explore this in detail in Section \ref{section:MSTWaccuracy}.

\subsection{Accuracy of evolved LO MSTW2008  distributions}\label{section:MSTWaccuracy}
\begin{figure}[h]
\begin{center}
\mbox{\epsfig{file=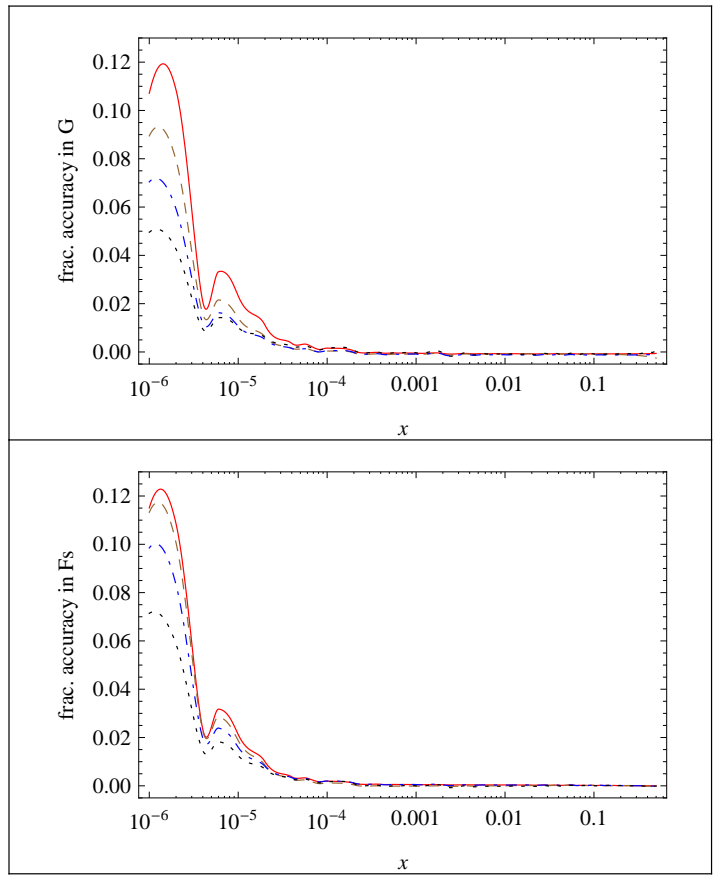
,width=4in%
,bbllx=0pt,bblly=0pt,bburx=317pt,bbury=370pt,clip=%
}}
\end{center}
\caption[]{
Fractional accuracy plots for LO MSTW \cite{MSTW1} gluon distributions $G(x,Q^2)=xg(x,Q^2)$ (upper plot) and $F_s(x,Q^2)$ distributions (lower plot), for $Q^2 = 5,\ 20,\ 100$ and $M_Z^2$ GeV$^2$, where the fractional accuracy is given by \eq{acc}.  The (red) curves are $Q^2 = 5$ GeV$^2$; the (brown) dashed curves are $Q^2 = 20$ GeV$^2$; the (blue) dot-dashed curves are $Q^2 = 100$ GeV$^2$; the (black) dotted curves are $M_Z^2$. The $x$ range covers all of the published LO MSTW $x$ data.
} 
\label{fig:GFsaccuracy}
\end{figure}

We now investigate quantitatively the accuracy of the {\em evolved}   LO  MSTW2008  distributions ($Q^2>Q_0^2$), introducing the fractional accuracy variable 
\ba
{\rm Fractional\ Accuracy} \equiv 1-f_{i,{\rm BDHM}}/f_{i,{\rm MSTW}}, \qquad i=1,2, 
\label{acc} 
\ea
 where $f_1=F_s,\ f_2=G$, with   BDHM denoting our LO analytic evaluations and MSTW denoting the published LO  MSTW2008 values \cite{MSTW1}.
We show in Fig. \ref{fig:GFsaccuracy} the fractional accuracy for the LO MSTW published distributions \cite{MSTW1} $G(x,Q^2)$ (upper figure) and $F_s(x,Q^2)$ (lower  figure) using the same  four $Q^2$ values  and legends used in Section \ref{section:singletMSTW} and Fig. \ref{fig:GFs}, i.e., the (red) curves are $Q^2 = 5$ GeV$^2$; the (brown) dashed curves are $Q^2 = 20$ GeV$^2$; the (blue) dot-dashed curves are $Q^2 = 100$ GeV$^2$; the (black) dotted curves are $M_Z^2$. Both the MSTW2008 $G$ and $F_s$ are in excellent agreement with our (much more numerically precise) calculations in the domain $x\ga 10^{-4}$, with a fractional accuracy of $\sim 0.1-0.5\%$. However, as is clearly seen in Fig. \ref{fig:GFs} for {\em both} $G$ and $F_s$ and for {\em all} $Q^2$, there is the same inaccuracy pattern in $x$, an increase of the fractional accuracy to $\sim 2\%$ down to $x\approx 8\times 10^{-6}$, followed by a dip at $x\approx 4\times 10^{-6}$, with a final rise to another maximum at $x\approx 2\times 10^{-6}$ whose fractional accuracy is $\sim 12\%$. These final inaccuracies at small $x$ are quite significant. Since the $x$ patterns are essentially independent of whether we are evaluating either $G$ or $F_s$, as well as being independent of $Q^2$, they suggest that the MSTW numerical program undergoes a significant structural change at some unique value of $x$, independent of $Q^2$, that seriously degrades their numerical output, leading to large errors at small $x$. The largest errors occur at the smallest $Q^2$; at $Q_0^2$ (not shown) the error is $\sim 12$---13 \%, and decreases monotonically to $\sim 4$---5 \% at the highest $Q^2$. As we will later see in Section \ref{section:CTEQ6L}, there is no such pattern in the LO CTEQ6L data \cite{CTEQ6.1}.

\section{ LO CTEQ6L singlet and gluon distributions}\label{section:CTEQ6L} 
As a second example of the application of our analytic decoupled solutions,  we will compare  our LO $x$-space gluon distribution $G(x,Q^2)=xg(x,Q^2)$ from \eq{newG} and our LO singlet distribution function $F_s(x,Q^2)$  from  \eq{newF}---using the published LO CTEQ6L \cite{CTEQ6.1} initial conditions at $Q_0^2=1.69 $ GeV$^2$---with the corresponding LO  CTEQ6L distributions \cite{CTEQ6.1}.  In order to insure continuity across the boundary $M_b^2$, we will first evolve from $Q_0^2=1.69$ GeV$^2$ (the CTEQ6L $Q_0^2$ value) to $M_b^2$ and use our evolved values of $G(x,M_c^2)$ and $F_s(x,M_c^2)$ for {\em new} starting values $G_0(x)$ and $F_{s0}(x)$, thus insuring continuity of $G$ and $F_s$ at the boundary where $n_f$ changes. We use the CTEQ6L values  $M_c=1.3$ GeV and $M_b=4.5$ GeV. We here use  a NLO version of $\alpha_s(Q^2)$, with $\alpha_s(M_Z^2)=0.118$, made continuous at $M_b$ and $M_c$, that was utilized in CTEQ6L (for details see Ref. \cite{CTEQ6.1}).

\begin{figure}[h]
\begin{center}
\mbox{\epsfig{file=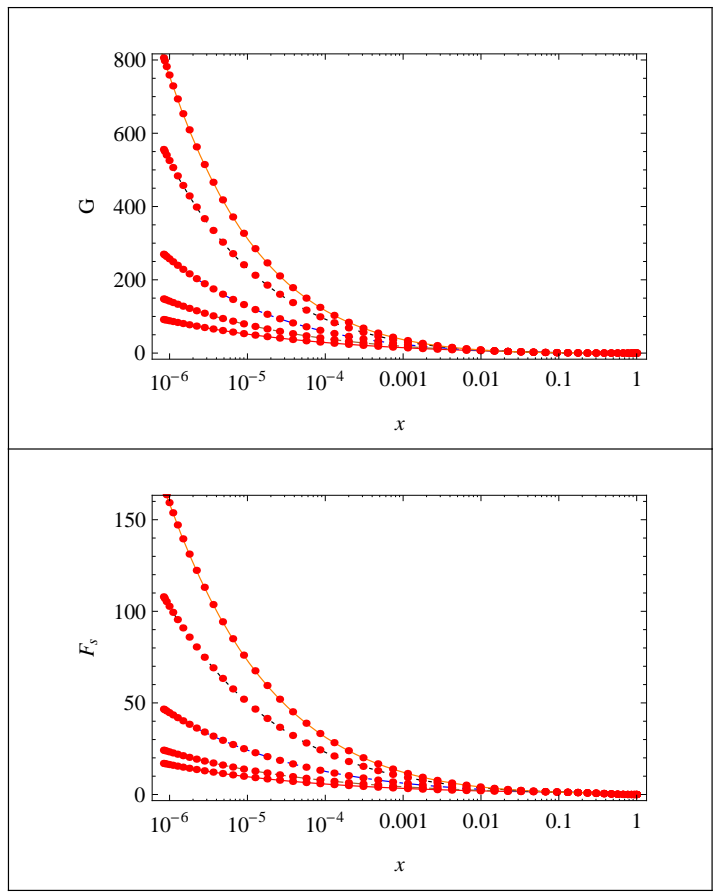
,width=4in%
,bbllx=0pt,bblly=0pt,bburx=317pt,bbury=380pt,clip=%
}}
\end{center}
\caption[]{
Plots for LO CTEQ6L \cite{CTEQ6.1} gluon distributions $G(x,Q^2)=xg(x,Q^2)$ (upper plot) and $F_s(x,Q^2)$ distributions (lower plot) vs. Bjorken $x$.
 The curves are  for $Q^2 = 10\ ,22,\ 90$,\ $1200$ and 
$M_Z^2$  GeV$^2$, bottom  to top. The (red) dots are our evolution results for LO $G(x,Q^2)$ from \eq{newG} and $F_s$ from \eq{newF} (converted to $x$-space) using the LO CTEQ6L values for $F_{s0}(x)$ and G$_0(x)$, where $Q^2_0=1.69$ GeV$^2$. The $x$ range in this Figure covers all of the published LO CTEQ6L $x$ data. 
} 
\label{fig:GFsCTEQ}
\end{figure}

\subsection{$G(x,Q^2)$ and $Fs(x,Q^2)$ for LO CTEQ6L }
In Fig. \ref{fig:GFsCTEQ} we show the Bjorken $x$-space results for LO $G(x,Q^2)=xg(x,Q^2)$ (upper figure) and LO $F_s(x,Q^2)$ (lower figure) vs. $x$, for 5 representative values of $Q^2$. The $x$-domain, $10^{-6}\le x \le 1$, is the complete region covered by the CTEQ group \cite{CTEQ6.1}.   The curves are the published CTEQ6L \cite{CTEQ6.1} LO  distributions. From bottom  to top; the (red) curve is for $Q^2 = 10$ GeV$^2$; the (brown) dashed curve is for $Q^2 = 22$ GeV$^2$;  the (blue) dot-dashed curve is for $Q^2 = 90$ GeV$^2$;   the (black) dotted curve is for $Q^2 =1200$ GeV$^2$; the (orange) curve is for $Q^2 = M_Z^2$. Since CTEQ6L \cite{CTEQ6.1} started evolution at  $Q_0^2=1.69$ GeV$^2$, we used $ F_{s0}$ and $ G_0$ constructed from their values at  $Q_0^2=1.69$ GeV$^2$ in  \eq{newG} and \eq{newF}. The (red) dots are our results for LO $G(x,Q^2)$ from \eq{newG} and $F_s(x,Q^2)$ from \eq{newF} converted to $x$-space, using  LO CTEQ6L values for $F_{s0}(x)$ and G$_0(x)$, evaluated using {\em Mathematica} \cite{Mathematica}.

For all $Q^2$ the agreement is excellent over the entire $x$ region, with a fractional accuracy of about $\pm \,5 \times 10^{-4}$, (completely consistent with  the 4 significant figures that are published)---for all $F_s$ and $G$ at the five  virtualities that we evaluated---with a minor and numerically unimportant exception of the lowest $x$ region of $F_s(x,Q^2=22)$, where there was an {\em offset} of $\approx 2\times 10^{-3}$.

\subsection{Accuracy of CTEQ6L {\em devolved} distributions}\label{section:MSTWacc}

In  Fig. \ref{fig:GFsCTEQ}, all of the distributions were for {\em evolutions} of $G$ and $F_s$ from the CTEQ6L $Q^2_0=1.69$ GeV$^2$ to {\em larger} $Q^2$. For another physics investigation, not relevant to this paper, we decided to compare LO starting distributions for MRSTW2008 and CTEQ6L at the MSTW2008 starting value of $Q_0^2=1$ GeV$^2$. Using $n_f=3$, we  {\em devolved} $G$ and $F_s$ from the CTEQ6L starting values at $Q^2_0=1.69$ GeV$^2$ {\em down} to $Q^2=1$ GeV$^2$, the MSTW2008 starting value for $Q_0^2$. 

\begin{figure}[h]
\begin{center}
\mbox{\epsfig{file=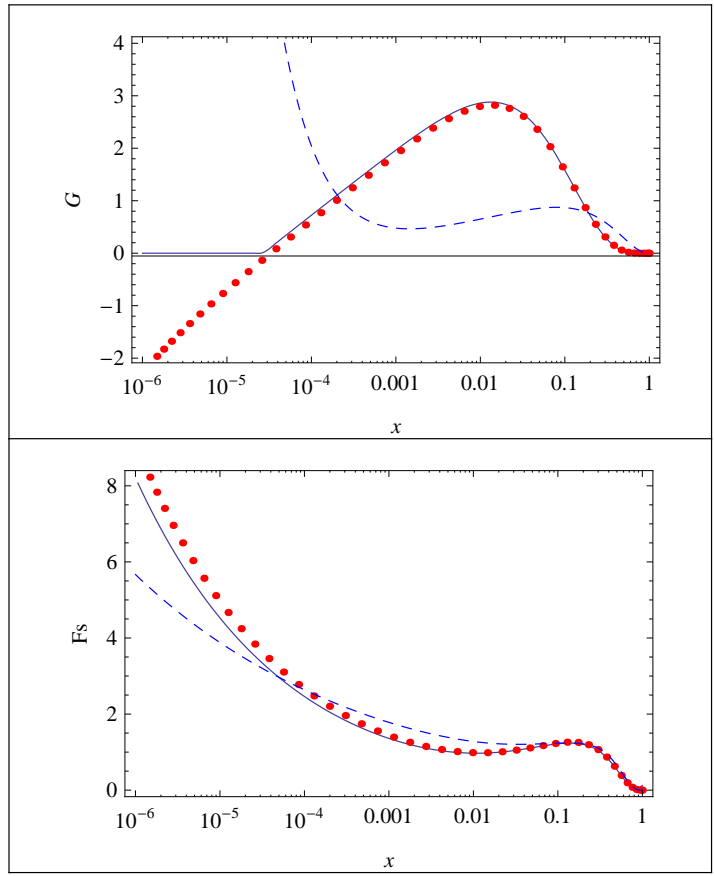
,width=4in%
,bbllx=0pt,bblly=0pt,bburx=317pt,bbury=370pt,clip=%
}}
\end{center}
\caption[]{
Plots of  gluon distributions $G(x,Q^2)=xg(x,Q^2)$ (upper plot) and $F_s(x,Q^2)$ distributions (lower plot) vs. Bjorken $x$, at the {\em devolved} value of $Q^2=1$ GeV$^2$. The (red) dots are our {\em devolution} results; the (black) solid curves are the published CTEQ6L results \cite{CTEQ6.1} and the (blue) dashed curves are the starting $F_{s0}(x)$ and $G_0(x)$ for MSTW2008 \cite{MSTW1}.
}
\label{fig:GFsCTEQdevolution}
\end{figure}

The results of this devolution are compared to the published CTEQ6L values \cite{CTEQ6.1} in Fig. \ref{fig:GFsCTEQdevolution} for $G$ (upper figure) and $F_s$ (lower figure). In all cases, when we refer to ``published CTEQ6L values'', we mean the results found on the Durham pdf generator web site;  see footnote %
\footnote{{ http://hepdata.cedar.ac.uk/pdf/pdf3.html}. The data used here were obtained in  August, 2010. We caution the reader that the web site format has been changed recently and that if one looks for CTEQ6L results for {\em any} $Q^2<M_c^2$, the site now returns the numerical values {\em for} $Q^2=M_c^2$; it functions normally for $Q^2\ge M_c^2$}.
The solid (black) curves are for CTEQ6L and the (red) dots are from \eq{newG} and \eq{newF}. In marked contrast to their evolution results, the CTEQ6L devolution results are numerically unstable, with $F_s$ being wrong by  $\approx 12\%$ at $x=10^{-6}$.  We also note that there are large disagreements with  their {\em devolved} $G(x)$ for small $x$.   Clearly, they have chosen to chop off their $G$ distribution at small $x$, i.e., to write $G(x)=0$ for small $x$, rather than allow it to become negative. The errors for both $G$ and $F_s$ become insignificant as $x$ approaches 1. It is clear that CTEQ encounters major problems with the numerical stability of their published results  for  $Q^2<Q^2_0$,  whereas they are completely accurate for  $Q^2>Q_0^2$.

For comparison, we also show in Fig.  \ref{fig:GFsCTEQdevolution} the  published MSTW2008 starting distributions \cite {MSTW1}  $G_0(x)$ and  $F_{s0}(x)$ at $Q^2_0=1$ GeV$^2$,  the dashed (blue) curves. We note  that the  LO gluon distributions of the two different collaborations, when evaluated at the {\em same} virtuality, $Q^2=1$ GeV$^2$,  bear little or no resemblance to each other, with the CTEQ6L gluon distribution going {\em negative} for $x\la 3\times 10^{-5}$. Although both singlet structure functions $F_s(x,Q^2=1)$ stay positive---as they must---Fig.  \ref{fig:GFsCTEQdevolution} shows that there are also large differences between the two singlet structure functions at low $x$.

\section{Overall numerical accuracy of analytical devolution and evolution}\label{section:accuracy}

\begin{figure}[h]
\begin{center}
\mbox{\epsfig{file=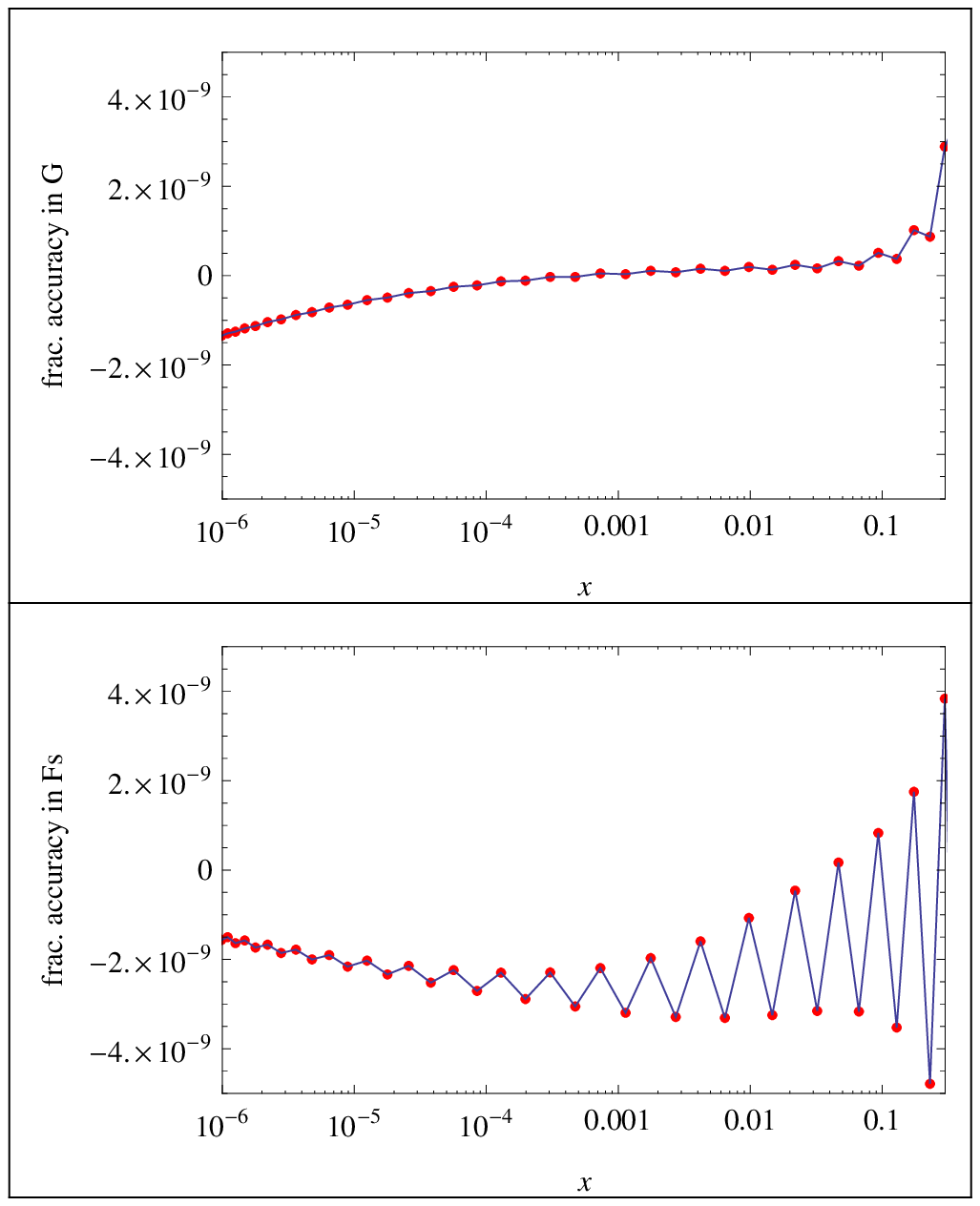
,width=4in%
,bbllx=0pt,bblly=0pt,bburx=317pt,bbury=370pt,clip=%
}}
\end{center}
\caption[]{
Fractional accuracy plots for our LO
 gluon distributions $G(x,Q^2)=xg(x,Q^2)$ (upper figure) from \eq{newG} and   $F_s(x,Q^2)$ distributions (lower figure) from \eq{newF}. These accuracy estimates resulted from {\em devolution} from $Q_0^2=1.69$ GeV$^2$ to $Q^2=1$ GeV$^2$, then using these results for {\em evolution} back to $Q^2=1.69$ GeV$^2$. The fractional value error estimates result from comparing the original values with the devolved-evolved ones.
}
\label{fig:BDHM_GFsaccuracy}
\end{figure}

As mentioned in Section \ref{section:MSTWacc}, we had {\em devolved} from $Q^2_0=1.69$ GeV$^2$ to $Q^2=1$ GeV$^2$, using the known CTEQ6L $G_0(x)$ and $F_{s0}(x)$ starting values.  To estimate the {\em overall} accuracy of our entire numerical procedure, we took our {\em devolved} distributions $G(x,Q^2=1)$ and $F_s(x,Q^2=1)$ and used {\em  them}  as starting values so that we could again {\em evolve} back to $Q^2=1.69$ GeV$^2$. Finally, we compared the evolved numerical results with the original $F_{s0}(x)$ and $G_0(x)$, the distributions that we started with at $Q^2=1.69$ GeV$^2$. An outline of our entire numerical procedure is given in the Appendix.

In Fig. \ref{fig:BDHM_GFsaccuracy}, we show the fractional accuracy of this ``round-trip'' comparison.  The upper figure
 is for $G$ and the lower figure is for $F_s$. The (red) dots are the ``round-trip'' fractional accuracies at discrete $x$-values  chosen to start and end this numerical exercise (corresponding to the transformed  zeroes of the Chebyshev polynomials that we discuss in the Appendix).  For the visual convenience of the reader, we have connected the dots. 

Where either $G$ and $F_s$  is significantly  large ($x\la 0.3$), we see that the ``round-trip'' error is $\la 4\times 10^{-9}$,  thus yielding an overall error estimate of $\la \pm \,2\times  10^{-9}$ for either evolution or devolution. Detailed causes for this error are discussed in  the Appendix. 

It is gratifying that the overall numerical uncertainty in our  LO  analytically decoupled solutions is small, thus furnishing us  not only with a new accurate and fast calculation tool for exploring the effects of the shapes of different starting value distributions, but also with  a  diagnostic tool for easily determining the numerical calculational  reliability of the already  published parton distribution functions that are currently in major use by the high energy physics community.

\section{Conclusions}

In conclusion, we have constructed {\em decoupled} analytical solutions for $F_s(x,Q^2)$ and $G(x,Q^2)$ from the coupled LO DGLAP equations, yielding  accurate   numerical results for {\em both} evolution and devolution of $O(10^{-9})$---a fast tool  to study the dependence on the shape of the starting distributions
 $F_{s0}(x)$ and $G_0(x)$, the boundary conditions at the starting value $Q_0^2$.  Similar procedures can be used for non-singlet distributions, allowing one to obtain {\em analytic} LO  solutions for {\em individual} quark distributions, as well as  for the gluon distribution; thus  avoiding the necessity for purely numerical solutions of the coupled DGLAP equations on a giant two-dimensional grid in ($x,\,Q^2$) space.  In essence, using  a program such as {\em Mathematica} \cite{Mathematica}, we can now define a parton distribution function for each quark and gluon and---after  inputting the desired $x$ and $Q^2$---evaluate it  accurately and rapidly (for a fast {\em Mathematica} program calculating  LO $F_s(x,Q^2)$ and $G(x,Q^2)$, see the Appendix).  

We have also used our  analytic solutions coupled with the MSTW2008 initial starting functions \cite{MSTW1} as a new and powerful diagnostic tool to study the numerical accuracy (the computational accuracy of their evolution code) of the LO MSTW2008 published distributions \cite{MSTW1}. For the small $x$-region, $x\la  10^{-4}$,   we discovered  a pattern of significant {\em numerical} (computational) errors for {\em both} $F_s$ and $G$,  ranging up to $\approx 12\%$ at the smallest $x$ values in the published MSTW2008 results \cite{MSTW1},   true for all $Q^2$.  

Applying the same new tools to  CTEQ6L, we found no errors (to their  accuracy of 4 significant figures) in either $F_s$ or $G$ values when they did {\em evolution} from $Q_0^2=1.69 $ GeV$^2$ to higher $Q^2$ values,  but significant errors---increasing with decreasing $x$---when they did {\em devolution} to smaller $Q^2$. In the future,  we intend to  evaluate $F_{s0}(x)$ and $G_{0}(x)$ in both LO and NLO, from a fit to small $x$ experimental data for the structure function $F_2^{\gamma p}(x,Q^2)$, in  order to obtain  (analytically) accurate values of $G(x,Q^2)$ {\em directly } tied to experiment, which are needed for the interpretation of experiments at the LHC.   
 
\section{Acknowledgments}
The authors would like to  thank the Aspen Center for Physics for its hospitality during the time parts of this work were done. P. Ha would like to thank Towson University Fisher College of Science and Mathematics for travel support.   D.W.M. receives support from DOE Grant No. DE-FG02-04ER41308.

\appendix
\section{}

We  outline here the actual calculation procedures necessary for  fast and accurate numerical evaluations of \eq{newF} and \eq{newG}. These calculations, although robust, require delicate choices as to the numerical techniques used in evaluating \eq{newF} and \eq{newG}.

As shown in Ref. \cite{inverseLaplace1}, if the function $g(s)$ goes to 0 at $\infty$ more  rapidly than $1/s$, then we can accurately approximate its inverse Laplace transform $G(v)$ by 
\ba
 G(v)&\approx& -{2\over  v} \sum^{N}_{i=1}{\rm Re}\left[\omega_ig\left({\alpha_i/ v}\right)\right],\label{Bromwich3}
\ea
where  $2N$ is the order of the approximation, $\omega_i$ and $\alpha_i$, $i=1,2,\ldots,2N$, are known complex numbers for a given $2N$, occurring in {\em complex conjugate pairs}. The actual numerical evaluation of \eq{Bromwich3} can be quite unstable if one doesn't utilize arbitrary accuracy arithmetic as discussed in Ref. \cite{inverseLaplace1}, since the weight functions $\omega_i$ become exceedingly large, even for modest $2N$,  and {\em oscillate} in sign \cite{inverseLaplace1}. The use of {\em Mathematica} (or similar programs, which also carry out arithmetical operations to arbitrary accuracy) makes this requirement easy to satisfy.

As shown in Ref. \cite{inverseLaplace1}, the inverse Laplace transform approximation to  $G(v)$ is {\em exact} if $G(v)$ is a polynomial in $v$ of order $4N-1$ or less.  For our purposes here, $G(v)$ in \eq{Bromwich3} is either the ${\hat F_s}(v,Q^2)$ or the ${\hat G}(v,Q^2)$ on the l.h.s. of \eq{newF} or \eq{newG}, whereas $g(s)$ in \eq{Bromwich3} is the surrogate for either 
$k_{ff}(s,\tau)f_0(s)+k_{fg}(s,\tau) g_0(s)$ found in the r.h.s. of  \eq{newF} or $k_{gg}(s,\tau)g_0(s)+k_{gf}(s,\tau) f_0(s)$ found in the r.h.s. of \eq{newG}.  Since we must evaluate $g(s)$ at {\em complex} values of $s$, this necessarily implies that we must evaluate $f_0(s)$ and $g_0(s)$---the Laplace transforms of $\hat F_{s0}(v)$ and $\hat G_{0}(v)$, respectively---at complex values of $s$.  As shown in Ref. \cite{inverseLaplace1}, to insure numerical accuracy  we must be able to evaluate $g(s)$ in \eq{Bromwich3} to arbitrary accuracy.  Thus we must know the Laplace transforms $f_0(s)$ and $g_0(s)$ {\em analytically} and {\em not}  just as {\em numerical integrations} of the form $\int_0^\infty\hat  F_{s0}
(v)e^{-vs}\,dv$.  The $k$'s, the coefficient functions needed, are known analytically; the potential problem is with  $f_0(s)$ and $g_0(s)$, the starting functions in Laplace space $s$. 

The starting  distributions functions normally  used are {\em not} of the type that have analytic Laplace transforms.  To get a sufficiently accurate numerical approximation to functions that {\em do} have analytic Laplace transforms is again a delicate numerical exercise.  We found that we could do it sufficiently accurately by using an interpolating polynomial of order n=49. Its 50 coefficients were determined by evaluating the original function at  50 points, distributed as the zeroes of  a 50$^{\rm th}$ order Chebyshev polynomial, found  in the interval $(-1,+1)$ and then linearly transformed to $ v$ space to lie in the interval $0\le v< 14 \ (1\ge x >0.83\times  10^{-6})$.  These points were chosen to try to minimize the maximum interpolation error. We note that even when using {\em Mathematica}, caution was needed in order  to obtain sufficient numerical accuracy  with a such a high order polynomial; it had to be evaluated using Horner's method (see Section 10.14 of Ref. \cite{Hildebrand}), since straight forward evaluation of such a high order polynomial will yield numerical nonsense.

Using $2N=38$ in \eq{Bromwich3}, we would have an {\em exact} result if either $\hat F_s(v,Q^2)$ in \eq{newF} or $\hat G(v,Q^2)$  \eq{newG} were a polynomial in $v$ of degree 75 or less;  see Ref. \cite{inverseLaplace1} for details.  In actual practice, by comparing the results for the value of $2N=38$ ---the value we used for our numerical evaluations--- with very much larger values of $2N$ that we used for estimates of the exact solutions, we found that the fractional accuracy of inversion for both $\hat F_s(v,Q^2)$ and $\hat G(v,Q^2)$ was $\approx \pm\, 1\times 10^{-11}$ for $v\ga 0.3$. Thus,  numerical inversion of the Laplace transform in either \eq{newF} or \eq{newG} contributes  essentially  nothing to our overall error of about $\pm\, 2\times 10^{-9}$, since it's  some 2 orders of magnitude  smaller.  We comment that the overall error  is essentially completely due to our numerical approximation of the starting functions and {\em not}  the subsequent Laplace transforms of them. Therefore, 
we could readily reduce this error by using more than 50 points in our numerical approximations of the starting distributions, but this would be at the expense of more computational time and was felt to be unnecessary. 

A typical time for computing the full $x$ distribution of either  $F_s(x)$ or $G(x)$ at an arbitrary $Q^2$---given the starting functions $F_{s0}(x)$ and $G_0(x)$ at $Q_0^2$---was about 15 seconds, basically proportional to the number of points in $x$ used in the numerical approximations of the starting functions and to the number $2N$ used in the Laplace inversion routine. Thus, for most applications, we could easily reduce this time to several seconds, at the expense of some (perhaps unneeded) accuracy. The computations in this paper were made on a home PC, a  Dell Model Studio XPS435MT, using an Intel 2.67 GHz 4 core i7 CPU, running 64 bit Windows Vista, and using {\em Mathematica7} \cite{Mathematica} in parallel mode. 

For a  very fast {\em Mathematica7} (.nb) program  that accurately calculates  all  LO MSTW2008 parton distribution functions, as well as $F_2^{\gamma p}(x)$ and $F_s(x)$ for any $Q^2$---using the LO MSTW starting values \cite {MSTW1} for $F_s(x)$, $G(x)$ at $Q_0^2=1$ GeV$^2$---send an email request to {\tt mblock@northwestern.edu} for {\tt MSTW.zip}.

\bibliography{gluonsPRD.bib}

\begin{thebibliography}{16}
\expandafter\ifx\csname natexlab\endcsname\relax\def\natexlab#1{#1}\fi
\expandafter\ifx\csname bibnamefont\endcsname\relax
  \def\bibnamefont#1{#1}\fi
\expandafter\ifx\csname bibfnamefont\endcsname\relax
  \def\bibfnamefont#1{#1}\fi
\expandafter\ifx\csname citenamefont\endcsname\relax
  \def\citenamefont#1{#1}\fi
\expandafter\ifx\csname url\endcsname\relax
  \def\url#1{\texttt{#1}}\fi
\expandafter\ifx\csname urlprefix\endcsname\relax\def\urlprefix{URL }\fi
\providecommand{\bibinfo}[2]{#2}
\providecommand{\eprint}[2][]{\url{#2}}

\bibitem[{\citenamefont{Gribov and Lipatov}(1972)}]{dglap1}
\bibinfo{author}{\bibfnamefont{V.~N.} \bibnamefont{Gribov}} \bibnamefont{and}
  \bibinfo{author}{\bibfnamefont{L.~N.} \bibnamefont{Lipatov}},
  \bibinfo{journal}{Sov. J. Nucl. Phys.} \textbf{\bibinfo{volume}{15}},
  \bibinfo{pages}{438} (\bibinfo{year}{1972}).

\bibitem[{\citenamefont{Altarelli and Parisi}(1977)}]{dglap2}
\bibinfo{author}{\bibfnamefont{G.}~\bibnamefont{Altarelli}} \bibnamefont{and}
  \bibinfo{author}{\bibfnamefont{G.}~\bibnamefont{Parisi}},
  \bibinfo{journal}{Nucl. Phys. B} \textbf{\bibinfo{volume}{126}},
  \bibinfo{pages}{298} (\bibinfo{year}{1977}).

\bibitem[{\citenamefont{Dokshitzer}(1977)}]{dglap3}
\bibinfo{author}{\bibfnamefont{Y.~L.} \bibnamefont{Dokshitzer}},
  \bibinfo{journal}{Sov. Phys. JETP} \textbf{\bibinfo{volume}{46}},
  \bibinfo{pages}{641} (\bibinfo{year}{1977}).

\bibitem[{\citenamefont{Block}(2010{\natexlab{a}})}]{inverseLaplace1}
\bibinfo{author}{\bibfnamefont{M.~M.} \bibnamefont{Block}},
  \bibinfo{journal}{Eur. Phys. J. C} \textbf{\bibinfo{volume}{65}},
  \bibinfo{pages}{1} (\bibinfo{year}{2010}{\natexlab{a}}),
  \eprint{arXiv:0907.4790 [hep-ph]}.

\bibitem[{\citenamefont{Block}(2010{\natexlab{b}})}]{inverseLaplace2}
\bibinfo{author}{\bibfnamefont{M.~M.} \bibnamefont{Block}},
  \bibinfo{journal}{Eur. Phys. J. C} \textbf{\bibinfo{volume}{68}},
  \bibinfo{pages}{683} (\bibinfo{year}{2010}{\natexlab{b}}),
  \eprint{arXiv:1004.3585 [hep-ph]}.

\bibitem[{\citenamefont{Martin et~al.}(2009)\citenamefont{Martin, Stirling,
  Thorne, and Watt}}]{MSTW1}
\bibinfo{author}{\bibfnamefont{A.~D.} \bibnamefont{Martin}},
  \bibinfo{author}{\bibfnamefont{W.~J.} \bibnamefont{Stirling}},
  \bibinfo{author}{\bibfnamefont{R.~S.} \bibnamefont{Thorne}},
  \bibnamefont{and} \bibinfo{author}{\bibfnamefont{G.}~\bibnamefont{Watt}},
  \bibinfo{journal}{Eur. Phys. J. C} \textbf{\bibinfo{volume}{63}},
  \bibinfo{pages}{189} (\bibinfo{year}{2009}), \eprint{arXiv:0901.0002
  [hep-ph]}.

\bibitem[{\citenamefont{Pumplin et~al.}(2002)}]{CTEQ6.1}
\bibinfo{author}{\bibfnamefont{J.}~\bibnamefont{Pumplin}} \bibnamefont{et~al.}
  (\bibinfo{collaboration}{CTEQ}), \bibinfo{journal}{J. High Energy Phys.}
  \textbf{\bibinfo{volume}{0207}}, \bibinfo{pages}{012} (\bibinfo{year}{2002}),
  \eprint{hep-ph/0201195}.

\bibitem[{\citenamefont{Stump et~al.}(2003)\citenamefont{Stump, Huston,
  Pumplin, Tung, Lai, Kuhlmann, and Owens}}]{CTEQ6L}
\bibinfo{author}{\bibfnamefont{D.}~\bibnamefont{Stump}},
  \bibinfo{author}{\bibfnamefont{J.}~\bibnamefont{Huston}},
  \bibinfo{author}{\bibfnamefont{J.}~\bibnamefont{Pumplin}},
  \bibinfo{author}{\bibfnamefont{W.}~\bibnamefont{Tung}},
  \bibinfo{author}{\bibfnamefont{H.}~\bibnamefont{Lai}},
  \bibinfo{author}{\bibfnamefont{S.}~\bibnamefont{Kuhlmann}}, \bibnamefont{and}
  \bibinfo{author}{\bibfnamefont{J.}~\bibnamefont{Owens}}, \bibinfo{journal}{J.
  High Energy Phys.} \textbf{\bibinfo{volume}{0310}}, \bibinfo{pages}{046}
  (\bibinfo{year}{2003}), \eprint{[hep-ph/0303013]}.

\bibitem[{\citenamefont{Tung et~al.}(2007)\citenamefont{Tung, Lai, Belyaev,
  Pumplin, Stump, and Yuan}}]{CTEQ6.5}
\bibinfo{author}{\bibfnamefont{W.~K.} \bibnamefont{Tung}},
  \bibinfo{author}{\bibfnamefont{H.~L.} \bibnamefont{Lai}},
  \bibinfo{author}{\bibfnamefont{A.}~\bibnamefont{Belyaev}},
  \bibinfo{author}{\bibfnamefont{J.}~\bibnamefont{Pumplin}},
  \bibinfo{author}{\bibfnamefont{D.}~\bibnamefont{Stump}}, \bibnamefont{and}
  \bibinfo{author}{\bibfnamefont{C.-P.} \bibnamefont{Yuan}},
  \bibinfo{journal}{J. High Energy Phys.} \textbf{\bibinfo{volume}{0702}},
  \bibinfo{pages}{053} (\bibinfo{year}{2007}), \eprint{hep-ph/0611254}.

\bibitem[{\citenamefont{Martin et~al.}(2002)\citenamefont{Martin, Roberts,
  Stirling, and Thorne}}]{MRST}
\bibinfo{author}{\bibfnamefont{A.~D.} \bibnamefont{Martin}},
  \bibinfo{author}{\bibfnamefont{R.~G.} \bibnamefont{Roberts}},
  \bibinfo{author}{\bibfnamefont{W.~J.} \bibnamefont{Stirling}},
  \bibnamefont{and} \bibinfo{author}{\bibfnamefont{R.~S.}
  \bibnamefont{Thorne}}, \bibinfo{journal}{Eur. Phys. J. C}
  \textbf{\bibinfo{volume}{23}}, \bibinfo{pages}{73} (\bibinfo{year}{2002}),
  \eprint{hep-ph/0110215}.

\bibitem[{\citenamefont{Martin et~al.}(2004)\citenamefont{Martin, Roberts,
  Stirling, and Thorne}}]{MRST4}
\bibinfo{author}{\bibfnamefont{A.~D.} \bibnamefont{Martin}},
  \bibinfo{author}{\bibfnamefont{R.~G.} \bibnamefont{Roberts}},
  \bibinfo{author}{\bibfnamefont{W.~J.} \bibnamefont{Stirling}},
  \bibnamefont{and} \bibinfo{author}{\bibfnamefont{R.~S.}
  \bibnamefont{Thorne}}, \bibinfo{journal}{Phys. Lett. B}
  \textbf{\bibinfo{volume}{604}}, \bibinfo{pages}{61} (\bibinfo{year}{2004}),
  \eprint{hep-ph/0410230}.

\bibitem[{\citenamefont{Block et~al.}(2008)\citenamefont{Block, Durand, and
  McKay}}]{bdm1}
\bibinfo{author}{\bibfnamefont{M.~M.} \bibnamefont{Block}},
  \bibinfo{author}{\bibfnamefont{L.}~\bibnamefont{Durand}}, \bibnamefont{and}
  \bibinfo{author}{\bibfnamefont{D.~W.} \bibnamefont{McKay}},
  \bibinfo{journal}{Phys. Rev. D} \textbf{\bibinfo{volume}{77}},
  \bibinfo{pages}{094003} (\bibinfo{year}{2008}), \eprint{arXiv:0710.3212
  [hep-ph]}.

\bibitem[{\citenamefont{Block et~al.}(2009)\citenamefont{Block, Durand, and
  McKay}}]{bdm2}
\bibinfo{author}{\bibfnamefont{M.~M.} \bibnamefont{Block}},
  \bibinfo{author}{\bibfnamefont{L.}~\bibnamefont{Durand}}, \bibnamefont{and}
  \bibinfo{author}{\bibfnamefont{D.~W.} \bibnamefont{McKay}},
  \bibinfo{journal}{Phys. Rev. D} \textbf{\bibinfo{volume}{79}},
  \bibinfo{pages}{014031} (\bibinfo{year}{2009}), \eprint{arXiv:0808.0201
  [hep-ph]}.

\bibitem[{\citenamefont{Block et~al.}(2010)\citenamefont{Block, Durand, Ha, and
  McKay}}]{bdhmNLO}
\bibinfo{author}{\bibfnamefont{M.~M.} \bibnamefont{Block}},
  \bibinfo{author}{\bibfnamefont{L.}~\bibnamefont{Durand}},
  \bibinfo{author}{\bibfnamefont{P.}~\bibnamefont{Ha}}, \bibnamefont{and}
  \bibinfo{author}{\bibfnamefont{D.~W.} \bibnamefont{McKay}},
  \bibinfo{journal}{Eur. Phys. J. C,} \textbf{\bibinfo{volume}{69}},
  \bibinfo{pages}{425} (\bibinfo{year}{2010}),
  \eprint{arXiv:1005.2556[hep-ph]}.

\bibitem[{Mat(2009)}]{Mathematica}
\bibinfo{journal}{{\em Mathematica} 7, a computing program from Wolfram
  Research, Inc., Champaign, IL, USA, www.wolfram.com}  (\bibinfo{year}{2009}).

\bibitem[{\citenamefont{Hildebrand}(1987)}]{Hildebrand}
\bibinfo{author}{\bibfnamefont{F.~B.} \bibnamefont{Hildebrand}},
  \bibinfo{journal}{Introduction to Numerical Analysis, Dover Publications, New
  York}  (\bibinfo{year}{1987}).

\end{thebibliography}

\end{document}